\documentclass[twocolumn,showpacs,aps, prb, floatfix, amsmath]{revtex4}
\usepackage{dcolumn}
\usepackage{bm}
\usepackage{graphicx}
\usepackage{bm}

\begin{document}
\title{Giant negative magnetoresistance and kinetic arrest of first-order ferrimagnetic-antiferomagnetic transition in Ge doped Mn$_2$Sb}
\author{Vikram Singh\footnote{Present address: School of Physics, Indian Institute of Science Education and Research, Thiruananthapuram, Kerala-695551, India}, R Rawat\footnote{E-mail : rrawat@csr.res.in}}
\affiliation{UGC-DAE Consortium for Scientific Research, University Campus, Khandwa Road, Indore-452001, India.}

\author{Pallavi Kushwaha}
\affiliation{CSIR-National Physical Laboratory, Dr. K S Krishnan Marg, New Delhi-110012, India.}

\begin{abstract}

Effect of Ge substitution on first order ferrimagnetic (FRI) - antiferromagnetic (AFM) transition in Mn$_2$Sb has been studied. It shows that transition temperature (T$_t$) can be tuned between 119~K - 271~K by substituting 2.5-10\% Ge at Sb site in Mn$_2$Sb. The variation of density of state at Fermi level N(E$_f$) with Ge substitution shows that dN(E)/dE is positive at E$_f$ in the AFM state.  With the application of magnetic field T$_t$ shifts to low temperature, which results in a giant negative magnetoresistance (MR) reaching a value of 70\% for 2.5\% substitution. Our results show that FRI to AFM transformation during cooling stops around 35 K, even though it remains incomplete. It along with non-monotonic variation of lower critical field, open loop in isothermal MR and increasing difference in zero field cooled warming (ZFCW) and field cooled warming (FCW) resistivity with increasing magnetic field shows that FRI to AFM transition is kinetically arrested in the case of 2.5\% Ge substitution.
  
\end{abstract}


\maketitle\section{Introduction}
Mn$_2$Sb is one of the oldest known ferrimagnet (FRI) which becomes paramagnetic above 550 K (T$_C$).\cite{Cullity2008,Wilkinson1957,Austin1963} With the substitution of transition element (e.g. Cr, V, Co, Cu, Zn) at Mn site, T$_C$ decreases, and at low temperatures a first order ferrimagnetic (FRI) to antiferromagnetic (AFM) transition  is observed.\cite{Bither1962, Galkin1970, Kanomata1984, Ryzhkovskii1990, Ryzhkovskii1992,Zhang2004, Kushwaha2008, Kushwaha2009PRB79} The FRI-AFM transition is also observed for As, Sn, Ge substitution for Sb,\cite{Zhang2004JPD, Vikram2014} which suggest that rather than magnetic ion dilution, the origin of this transition could be structural or electronic or both. In fact, both resistivity and lattice parameters are reported to show large variation across FRI-AFM transition. Due to strong coupling between spin, charge and lattice, these systems show many functional properties like giant magnetoresistance (MR), magnetostriction, magnetocaloric effect (MCE) etc. \cite{Zhang2004JPD,Vikram2014,Wijingaard1992, ZhangPRB2003, Zhang2004, Caron2013, Bartashevich2002, Zhang2008}
	
	 A survey of the phase diagrams of Mn$_2$Sb with dopant concentration shows that FRI-AFM transition temperature ($T_t$) is either absent or higher than 100 K.\cite{Bither1962, Kanomata1984, Bartashevich2002} Even though, there appears to be some discrepancies in T$_t$ vs. dopant concentration for most of the dopant, the results from independent groups are more or less consistent for Cr substitution at Mn site and Ge substitution at Sb site.\cite{Bither1962, Galkin1970, Kanomata1984, Ryzhkovskii1990, Ryzhkovskii1992, Zhang2004, Kushwaha2008, Kushwaha2009PRB79, Vikram2014, ZhangPRB2004, ZhangPRB2004E,Jarrett1961,Swoboda1960, Cloud1960, Wijingaard1992, Caron2013,ZhangPRB2003,Engelhardt1999, Zhang2004JPD, Shimada2013, Koyama2013, Wakamori2016} Both, leads to decrease in one conduction electron per atom substitution as well as decrease in unit cell volume and T$_t$ increases from around 75 K to room temperature. However, most of above cited work are related to Cr substitution. Magnetic measurements on Mn$_{1.977}$Cr$_{0.023}$Sb (with the lowest T$_t$) by Darnell et al.\cite{Darnell1963} shows that magnetization remains significantly higher well below T$_t$. This could be an indication of co-existing AFM and FRI phases below T$_t$. Phase co-existence at low temperature has also been reported in Co doped Mn$_2$Sb under pressure or magnetic field.\cite{Bartashevich2002, Kushwaha2008,Orihashi2013} In an attempt to shift the T$_t$ of Mn$_{1.85}$Co$_{0.15}$Sb to near zero K with  magnetic field, Kushwaha et al.\cite{Kushwaha2008} showed that this phase co-existence is a result of the kinetic arrest of first order FRI-AFM transition. Subsequently, similar results has also been reported for some other dopant \cite{Orihashi2013, Tekgul2017, Mitsui2018} e.g. Mitsui et al.\cite{Mitsui2018} showed that relaxation time for FRI to AFM transition in Cr doped Mn$_2$Sb increases by several orders at low temperature. However, in the case of Ge substituted Mn$_2$Sb, Shimada et al.\cite{Shimada2013} reported the absence of kinetic arrest. This observation is made for compositions with more than 5\% Ge substitution for which, T$_t$ is reported to be higher than 100 K. It is significantly larger than the temperatures where, kinetic arrest of transition is reported to dominate in Co doped Mn$_2$Sb. Therefore, to verify the presence of kinetic arrest in Ge doped Mn$_2$Sb system, investigations on composition with lower T$_t$ are required.  
	 
	Most of the existing reports on Ge substitution are limited to few compositions with $x$ $\ge$ 0.05. The composition around $x$ = 0.1 has been studied due to its near room temperature $T_t$ and associated MR and magnetocaloric effect.\cite{Zhang2004JPD, Vikram2014} On the other hand we could not find any report on compositions with $x <$ 0.05. Here we present our systematic investigation of FRI-AFM transition in Ge substituted Mn$_2$Sb. Transition temperature (T$_t$) is tuned between 119 K to 271 K by substituting 2.5-10\% Ge at Sb site. For 2.5\% substitution a detailed resistivity ($\rho$)/MR study down to 5 K and up to 16 Tesla magnetic field is carried out to verify the presence of kinetic arrest.

\maketitle\section{Sample preparation and experimental details}

\begin{figure}[b]
	\begin{center}
	\includegraphics[width=8 cm]{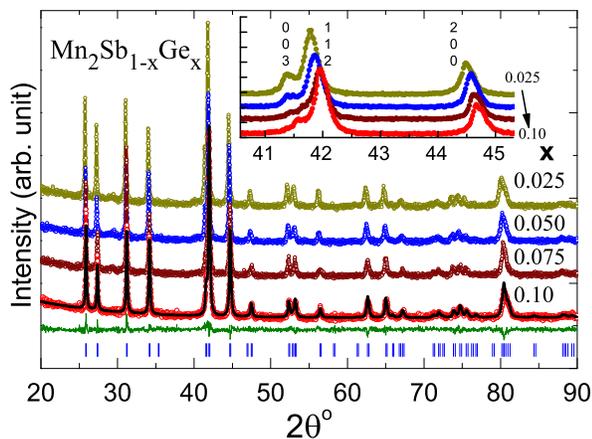}
	\end{center}
	\caption{(color online) Powder X-ray diffraction pattern of Mn$_2$Sb$_{1-x}$Ge$_x$ for $x$ = 0.025, 0.05, 0.075 and 0.10. Rietveld refined curve and its difference from the measured curve is shown as black and green line curves, respectively for $x$ = 0.1. The inset highlights the shift in [003], [112] and [200] peak position to higher angle with x.}
	\label{Figure1}
\end{figure}
	
	Samples of Mn$_2$Sb$_{1-x}$Ge$_x$ with $x$ = 0.025, 0.05, 0.075  and 0.10 are prepared by arc melting the stoichiometric amount of constituent elements of purity better than $ 99.98 $ \% under high purity argon gas atmosphere. To determine phase purity and structural parameters, powder X-ray diffraction (XRD) measurements at room temperature are performed using Cu K$_\alpha$ radiation. Figure 1 shows the measured XRD patterns. The peak positions shift to higher angle with increase in $x$, as highlighted in the inset. These patterns are fitted with Cu$_2$Sb-type tetragonal crystal structure with space group P$4$/nmm using the FullProf Reitvel refinement program.\cite{Rodriguez1993} A representative fit is shown as a line curve in figure 1 for $x$ = 0.1. The small un-indexed peak around $2\theta$ = 30$^0$ indicates small amount of MnSb impurity phase.\cite{Wilkinson1957}   The unit cell parameters `a', `c' and volume `V' obtained from this analysis are tabulated in Table-1 and these decreases monotonically with increases in $x$. The resistivity ($\rho$) in the temperature range 2-300 K is measured with a home-made setup along with 8-Tesla and 16-Tesla superconducting magnet systems from M/s. Oxford Instruments, U.K. Resistivity measurements in the presence of magnetic field are carried out in longitudinal geometry and magnetoresistance (MR) is defined as 100*$ \left\lbrace \rho (H, T)- \rho (0, T) \right\rbrace / \rho(0, T)$. Heat capacity (C$_P$) measurements in the temperature range 5-300 K are performed using a home-made setup based on semi-adiabatic heat pulse method.\cite{Rawat2001, Bag2014} Magnetization is measured using 14-Tesla vibrating sample magnetometer from M/s. Quantum Design, USA at TIFR, Mumbai.
	
\maketitle\section{Results and Discussion}

The temperature dependence of magnetization in the presence of 0.1 Tesla magnetic field is shown in figure 2 for $x$ =  0.025, 0.05 and 0.10. These measurements are carried out during cooling (FCC) and subsequent warming (FCW). A sharp decrease in magnetization on cooling correspond to the FRI to AFM transformation. Hysteresis around the transition region, which is distinctly visible for $x$= 0.025, indicates first order nature of the transition. Transition temperature (T$_t$) is taken as average of temperatures T* and T**, at which temperature derivative of magnetization shows a maxima during cooling and warming, respectively. As summarized in Table 1, T$_t$ increases from 119 K to 271 K with increase in $x$ from 0.025 to 0.1 and these are found to be consistent with existing data for x$\geq$0.05.\cite{Shimada2013,Koyama2013} A hump like feature around 220~K for x=0.025  can be attributed to spin-reorientation transition, at which direction of magnetic moment in the FRI state changes from c-axis (at room temperature) to ab-plane (at low temperatures).

\begin{figure}[h]
	\begin{center}
	\includegraphics[width=8 cm]{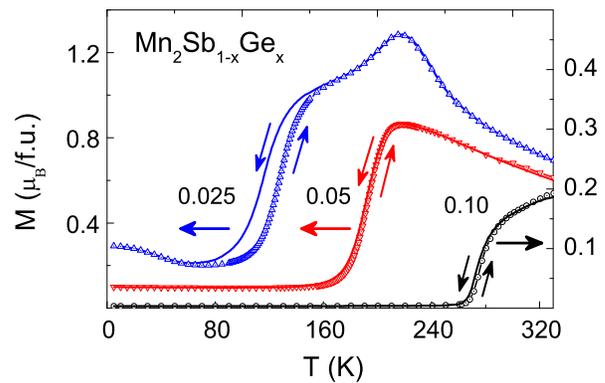}
	\end{center}
	\caption{(color online) Temperature dependence of magnetization (M) measured in the presence of 0.1 Tesla magnetic field during cooling (FCC) and subsequent warming (FCW) for Mn$_2$Sb$_{1-x}$Ge$_x$. Y-scale on the left hand side corresponds to $x$=0.025 and 0.05, and on the right hand side corresponds to $x$ = 0.10.}
	\label{Figure2}
\end{figure}
	 
 \begin{figure}[b]
	\begin{center}
		\includegraphics[width=8cm]{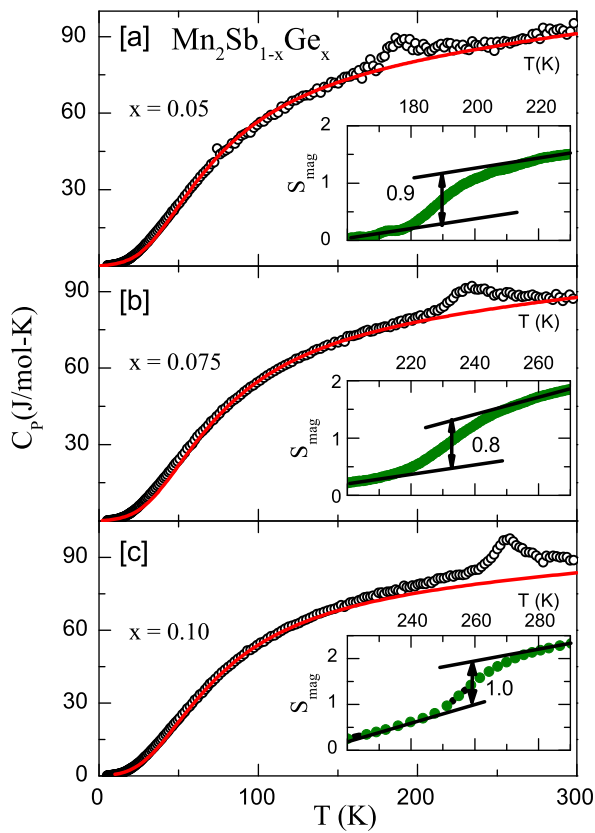}
	\end{center}
	\caption{(color online) The temperature dependence of measured heat capacity C$_P$  (black circle) and simulated nonmagnetic contribution (red line) for Mn$_2$Sb$_{1-x}$Ge$_x$ \textbf{[a]} $x$ = 0.05, \textbf{[b]} $x$ =0.075 and \textbf{[c]} $x$= 0.10. The insets show temperature variation of magnetic entropy (S$_{mag}$) around the transition region. See text for details.}
	\label{Figure3}
\end{figure}

\begin{figure}[b]
	\begin{center}
		\includegraphics[width=8cm]{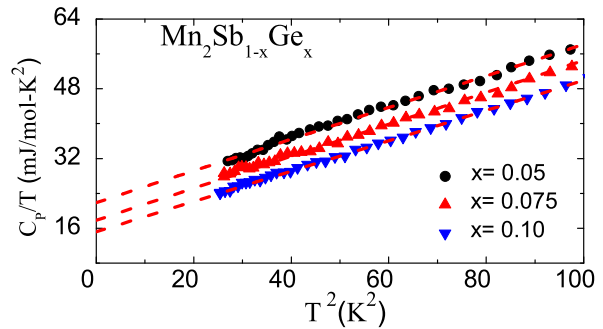}
	\end{center}
	\caption{(color online) Low temperature behavior of measured heat capacity ($C_P$) in the form of  C$_P$/T versus T$^2$ for Mn$_2$Sb$_{1-x}$Ge$_x$ (open symbol). Red dashed lines are extrapolated linear fit for the corresponding curves.}
	\label{Figure4}
\end{figure}

\begin{figure}[b]
	\begin{center}
		\includegraphics[width=8cm]{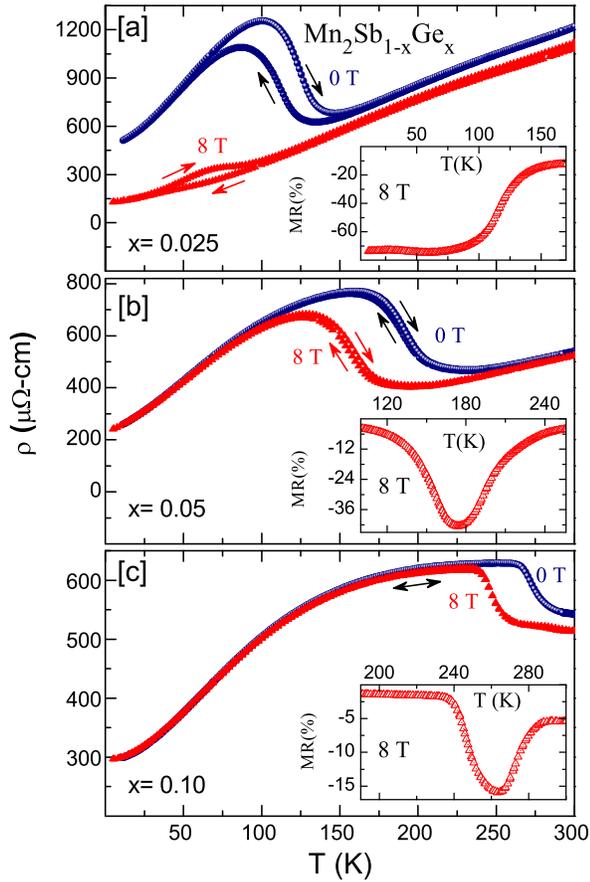}
	\end{center}
	\caption{(color online)Temperature dependence of resistivity ($\rho$) in 0 and 8 Tesla magnetic field measured during cooling and subsequent warming for Mn$_2$Sb$_{1-x}$Ge$_x$ \textbf{[a]}$x$ = 0.025, \textbf{[b]} $x$ =0.05 and \textbf{[c]} $x$= 0.10. Inset highlights magnetoresistance (MR) for 8 Tesla magnetic field calculated from the respective curves measured during warming.}
	\label{Figure5}
\end{figure}

	\begin{table*}
		\caption{\label{arttype} Variation of unit cell parameter `a', `c' and volume `V' , first order transition temperature (T$_t$), Debye temperature obtained from low temperature ($\theta_{D\_LT}$)/high temperature ($\theta_{D\_HT}$) heat capacity data, coefficient of electronic contribution to specific heat ($\gamma$), density of state at the Fermi level N(E$_f$) and entropy of AFM-FRI transition ($\Delta$S$_{mag}$) for Mn$_2$Sb$_{1-x}$Ge$_x$}.
		\begin{ruledtabular}	
			\begin{tabular}{ccccccccc}
				Mn$_2$Sb$_{1-x}$Ge$_x$ &a&c&V &T$_t$ &$\theta_{D\_HT}/\theta_{D\_LT}$  &$\gamma$ &N(E$_f$) &$\Delta S_{mag}$
				\\
				
				$x$&(\AA)&(\AA)&(\AA$^3$)&(K) &(K) &(mJ/mol-K$^2$)  & 10$^{24}$ (eV.mol)$^{-1}$ &(J/mol-K)\\
				\\
				0.025&4.068&6.539&108.2&118.9&---&---&---&---\\
				0.050&4.061&6.534&107.7&191.0&295/252&22&56&1.5\\
				0.075&4.057&6.516&107.3&245.0&296/252&18&46&1.9\\
				0.100&4.051&6.513&106.9&271.4&298/256&15&38&2.4\\
			\end{tabular}
		\end{ruledtabular}
	\end{table*}	
	 
	\begin{figure}[h]
		\begin{center}
		\includegraphics[width=\linewidth]{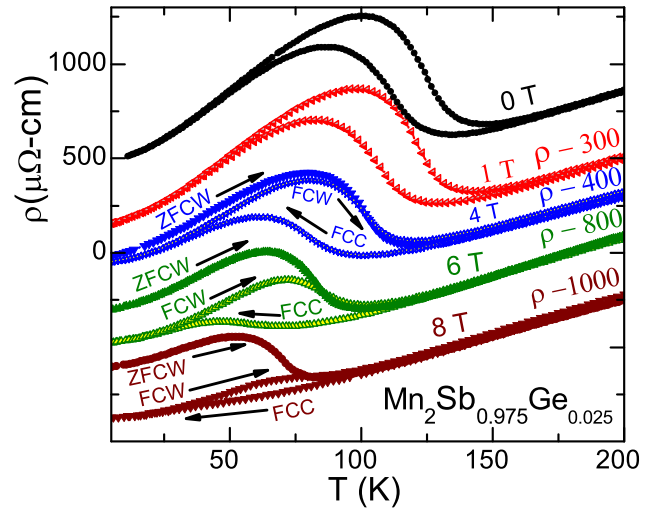}
		\end{center}
		\caption{(color online)Temperature dependence of resistivity ($\rho$) of Mn$_2$Sb$_{0.975}$Ge$_{0.025}$ in the presence of labeled magnetic field. The magnetic field is applied isothermally at 5 K after zero field cooling and $\rho$ is measured during warming (ZFCW), cooling (FCC) and subsequent warming (FCW). All $\rho$ curves, except for 0 T, are translated downward for the sake of clarity.}
		\label{Figure6}
	\end{figure}
 
	\begin{figure*}[t]
	\begin{center}
		\includegraphics[width=16cm]{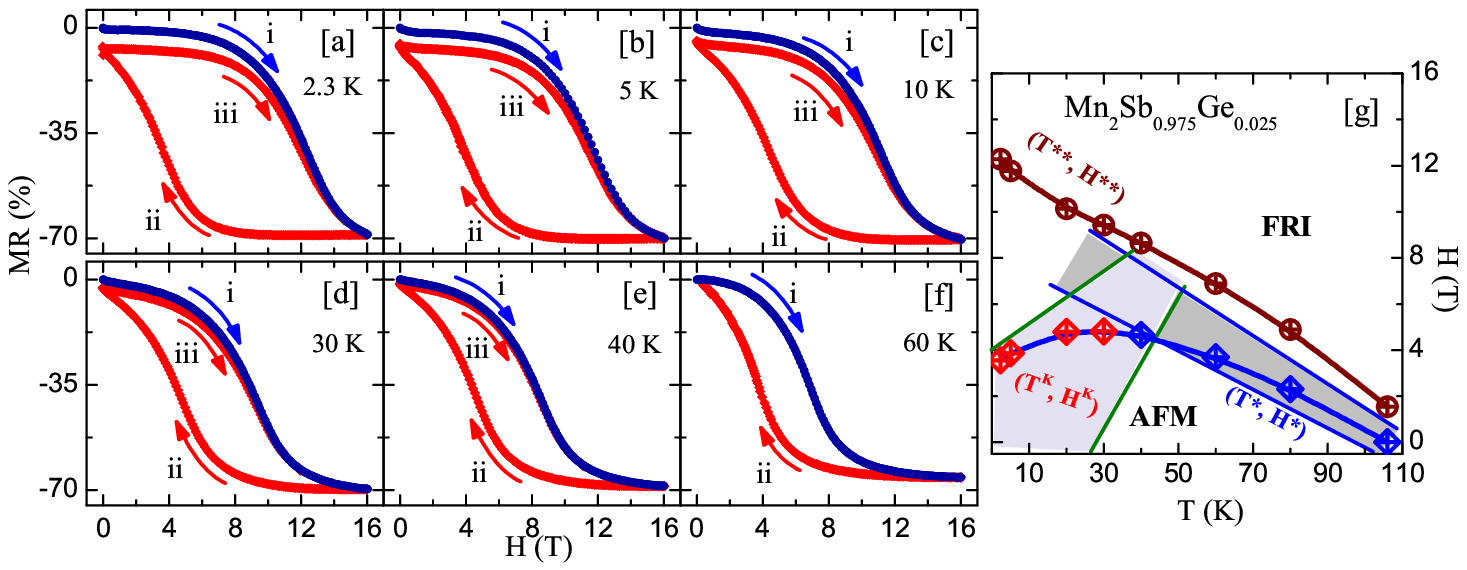}
	\end{center}
	\caption{(color online) \textbf{[a]-[f]}Isothermal magnetoresistance (MR) measured up to 16 Tesla magnetic field. For these measurements sample is cooled in zero magnetic field to the measurement temperature and then MR is measured with varying magnetic field 0 to 16 Tesla (curve-i), 16 to 0 Tesla (curve-ii) and 0 to 16 Tesla (curve-iii). \textbf{[g]} H-T phase diagram obtained from the isothermal MR measurement along with schematic of Kinetic arrest band and supercooling band.}
	\label{Figure7}
	\end{figure*}	

The temperature dependence of specific heat (C$_P$) measured during warming is shown in figure 3 for $x$ = 0.05, 0.075 and 0.1. The peak around T** corresponds to AFM-FRI transition. The line curves in the figure 3 are simulated non-magnetic contribution, which is taken as a combination of Debye term (lattice) and a linear term (electronic and anharmonic contributions). The Debye temperatures  for all the three compositions are found to be in the range 290~K to 300~K and these are tabulated as $\theta_{D\_HT}$ in the Table 1. The entropy change associated with the transition (S$_{mag}$) obtained from the difference of measured and simulated curves is shown in the insets of figure 3. The discontinuous jump around T$_t$ is found to be around 1 J/mol-K which corresponds to latent heat of 240 J/mol. Total magnetic entropy change   across the transition increases almost linearly from 1.5 J/mol-K to 2.4 J/mol-K with increase in $x$ from 0.05 to 0.1. Galkin et al.\cite{Galkin1970} estimated the entropy of the transition for $x$ = 0.12-0.20 using magnetization change at the transition and rate of change of critical field. In their study, the compositions with x=0.12 and 0.16 show multiple transitions and entropy associated with low temperature transition is found vary from  1.3 to 0.7 J/mol-K. The reported value for $x$=0.12 is close to our measured discontinuous change in S$_{mag}$ for $x$=0.10, which has nearly similar T$_t$. Similar values (1.3 J/mole K) has been reported by Engelhardt et al.\cite{Engelhardt1999} for Cr doped Mn$_2$Sb. 
	
	The low temperature behavior of C$_P$ in the form of  C$_P$/T vs T$^2$ is presented in figure 4. It shows a linear behavior ($\gamma$ + $\beta$T) with similar slope for all the three compositions. The Debye temperature obtained from the slope ($\beta$) of these curves is found to be around 250 K, which is tabulated as $\theta_{D\_LT}$ in Table 1. It is about 13\% less than that obtained from heat capacity data at higher temperatures. Such differences can arise due to non-linearity in phonon-dispersion curve. The intercept on Y-axis of extrapolated linear curves i.e. $\gamma$ (the coefficient of electronic contribution to C$_P$) decreases with increase in $x$. The $\gamma$ value for $x$= 0.10 composition is in good agreement  14 mJ/mol-K$^2$ for $x$=0.12 by Zhang and Zhang \cite{Zhang2004JPD}. Our study of $\gamma$ variation with Ge substitution shows that integrated density of state at the Fermi level N(E$_f$) decreases with Ge substitution in the AFM state. Using the relation $\gamma$ = $\pi^2 k_B^2$N(E$_f$)/3 we find that  N(E$_f$) varies from 5.6$x$10$^{24}$ to 3.8$x$10$^{24}$ (eV.mol)$^{-1}$. In the rigid band model it indicates that dN(E$_f$)/dE is positive, as Ge has one electron less than Sb in its outermost p-orbital.

	Figure 5 shows the resistivity measured in the absence and in the presence of 8 Tesla magnetic field during field cooling (FCC) and field cooled warming (FCW) for $x$ = 0.025, 0.05 and 0.10. The FRI to AFM transition during cooling results in resistivity increase. The resistivity behaviour in the AFM state shows a tendency to saturation at high temperature e.g. see figure 5(c) for $x$ = 0.10 with T$_t$ around 270 K.  The temperature dependence of resistivity has striking similarity with the Cr doped Mn$_2$Sb.\cite{Caron2013,Bierstedt1963} The tendency to saturation at high temperature  has been commonly observed in Lave phase compound  when resistivity approaches a limiting value. It along with rapid rise in resistivity at low temperature in the AFM state suggest that Fermi level may be lying below the peak in N(E) vs. E curve which also appears to be consistent with $\gamma$ variation with $x$. In the FRI state the tendency to saturation appears to be quite weak as in this case Fermi level is expected to be within peak region. With the application of magnetic field T$_t$ shifts to lower temperature. For $x$ = 0.05 and 0.10, it results in large negative magnetoresistance with a maximum magnitude around T$_t$. In the case of $x$ = 0.025, MR reaches a value of 70\% below T$_t$, which remains nearly constant down to the lowest temperature of measurement. Resistivity measured in the presence of 8 Tesla magnetic field for this sample shows much smaller change in resistivity around FRI-AFM transition, which suggest presence of untransformed FRI phase fraction at low temperatures. This feature is explored further under various constant magnetic field as shown in figure 6. Here, resistivity is measured during zero field cooled warming (ZFCW), FCC and FCW under labeled constant magnetic field.  All the resistivity curves, except for 0~T, are shifted vertically downward for the sake of clarity. The decrease in T$_t$ with the application of magnetic field is evident from these figures. For magnetic field value $\geq$ 4 Tesla, difference between ZFCW and FCC curve at low temperature can be seen, which increases for higher magnetic field. The lower value of resistivity for FCW in comparison to ZFCW indicate the presence of higher FRI phase fraction at 5 K for same field value. Another noticeable feature is the absence of FRI to AFM transformation below 30 K (hysteresis is closed below 30 K) irrespective of transformed FRI phase fraction. Similar resistivity behavior has been reported for Co doped Mn$_2$Sb,\cite{Kushwaha2008} Pd doped FeRh,\cite{Kushwaha2009, Saha2018} Al or Ru doped CeFe$_2$,\cite{Manekar2001, Sokhey2004} Se doped CoS$_2$.\cite{Mishra2016} In these system it has been attributed to kinetic arrest of the first order transition i.e. at low temperatures due to slow dynamics of the transition, the transformation from high temperature state to low temperature state is hindered on experimental time scale.\cite{Manekar2001, Kumar2006, Roy2008, Roy2013} It is known to result in open loop in isothermal MR\cite{Manekar2001, Roy2008} and non-monotonic variation of critical field with temperature.\cite{Rawat2007}

	To verify it, isothermal MR measurements up to 16 Tesla magnetic field are shown in fig 7 (a)-(f) for x = 0.025. For these measurements, sample is cooled in zero field from 300 K to the measurement temperature. It shows giant negative magnetoresistance of about 70\% due to field induced AFM to FRI transition. Another obvious feature at low temperature is that the MR (or resistance) at zero magnetic field before and after field cycling up to 16 Tesla does not have same value at low temperatures($<$30 K). Here, the virgin curve (curve-i) lies outside the envelope curve (curve-ii and iii), which indicates that field induced FRI phase do not transform back completely to its initial AFM state. The difference between virgin and envelope increases at lower temperatures.
	 
	The variation of critical field required for AFM-FRI transition with temperature is shown in the form of H-T phase diagram in figure 7(g). It shows that above 50 K, both lower (H* i.e. field required for FRI-AFM transition) and upper (H** i.e. field required for AFM-FRI transition) critical field increases with decrease in temperature, which is expected for a first order transition from low-temperature low-magnetization state to high-temperature high-magnetization state. However, below 40 K lower critical field (H*) starts decreasing with decrease in temperature. This phase diagram is qualitatively similar to that reported for Co doped Mn$_2$Sb\cite{Kushwaha2008} where, non-monotonic behavior of H* with temperature has been attributed to interplay between kinetics and thermodynamics of the transition.\cite{Rawat2007} The positive slope for lower critical field is dictated by the kinetics of the transition and it is represented as kinetic arrest band (H$_K$, T$_K$).\cite{Manekar2001, Kumar2006, Rawat2007, Kushwaha2009, Mishra2016, Saha2018} The left side of (H$_K$, T$_K$) band is the region where, kinetics of the transition dominates and FRI phase in this region will not be able to transform back to AFM state due to slow dynamics. Therefore a field induced FRI phase in this temperature region will not transform back to AFM state with decrease in magnetic field resulting in open loop in MR. Similarly during cooling if the (H$_K$, T$_K$) band is crossed before crossing the (H*, T*) then that region will remain in FRI state, which gives rise to difference in ZFCW and FCW. A survey of reported transition temperature variation with various dopant concentration in Mn$_2$Sb show that AFM-FRI tranformation takes place only above 50 K.\cite{Bither1962, Kanomata1984, Bartashevich2002, Darnell1963, Shimada2013} In the light of magnetic field tuning of T$_t$, it can be inferred that absence of first order transition in doped Mn$_2$Sb below 50 K could be related to kinetic arrest.

\maketitle\section{Conclusions}
		
		Our study of first order AFM-FRI transition in Mn$_2$Sb$_{1-x}$Ge$_x$ for $x$ = 0.025 - 0.1 shows increase in T$_t$ with $x$ and it varies almost linearly with unit cell volume. The decrease in electronic contribution to heat capacity owith $x$ suggest that Fermi level is lying below the peak but not at minima in the AFM state. As a result resistivity shows a rapid rise at low temperature followed by a saturation tendency at higher temperatures in the AFM state. Application of magnetic field in the AFM state induces FRI state, which results in giant negative magnetoresistance, reaching a value of more than 70\% for $x$ =0.025. Isothermal MR studies under high magnetic field in x=0.025 shows an open loop and non-monotonic variation of critical field required for FRI to AFM transition. These features confirm the presence of kinetic arrest below 30 K in Ge doped Mn$_2$Sb system, which also explains the absence of transition below 50 K with composition variation.
		
\maketitle\section{Acknowledgement}
Mukul Gupta and Layanta Behera are acknowledged for XRD measurements. PK acknowledges DST-SERB, India for support through Ramanujan Fellowship.


{}

\end{document}